\documentclass{IEEEtran}

\usepackage{cite}
\usepackage{amsmath,amssymb,amsfonts}
\usepackage{algorithmic}
\usepackage{graphicx}
\usepackage{textcomp}
\usepackage{subfigure}
\usepackage{authblk}
\def\BibTeX{{\rm B\kern-.05em{\sc i\kern-.025em b}\kern-.08em
    T\kern-.1667em\lower.7ex\hbox{E}\kern-.125emX}}
\begin{document}
\title{Joint Timing Offset and Channel Estimation for Multi-user UFMC Uplink}
\author[1,2]{Yicheng Xu}
\author[2,3]{*Hongyun Chu}
\author[3]{Xiaodong Wang, \textit{Fellow, IEEE}}
\affil[1]{National Mobile Communication Research Laboratory, Southeast University, Nanjing, 211189, China}
\affil[2]{Wireless Energy and Information Transmission Lab, Shenzhen Institutes of Advanced Technology\authorcr Chinese Academy of Sciences, Shenzhen, 518055, China}
\affil[3]{Department of Electrical Engineering, Columbia University, New York 10027, USA}






\maketitle

\begin{abstract}
Universal filtered multi-carrier (UFMC), which groups and filters subcarriers before transmission, is a potential multi-carrier modulation technique investigated for the emerging Machine-Type Communications (MTC). Considering the relaxed timing synchronization requirement of UFMC, we design a novel joint timing synchronization and channel estimation method for multi-user UFMC uplink transmission. 
Aiming at reducing overhead for higher system performance, the joint estimation problem is formulated 
using atomic norm minimization that enhances the sparsity of timing offset in the continuous frequency domain. Simulation results show that the proposed method can achieve considerable performance gain, as compared with its counterparts.
\end{abstract}


\section{Introduction}
UFMC, also known as universal-filtered orthogonal frequency-division multiplexing (UF-OFDM), does not need the cyclic prefix (CP) and has better spectral property compared with OFDM\cite{c39,c40}. In a UFMC system, the whole bandwidth is divided into several sub-bands each consisting of a group of sub-carriers, and each sub-band employs an FIR filter to suppress the spectral side-lobe levels and achieve higher robustness. 
The suppressed side-lobe significantly reduces the inter-block interferences caused by synchronization errors, and the centralized signal power decreases the missing information under slight time misalignment, which relaxes the synchronization requirement of UFMC and makes it possible more flexible resource allocations than in OFDM.

Although UFMC has lower sensitivity to timing offset (TO) compared with OFDM, time synchronization is still an important and open problem for UFMC-based systems \cite{c10,c11}. 
Generally, in uplink transmissions, a random access initialization, including the user equipment (UE) transmitting time alignment, should be processed before data exchanging between the base station (BS) and UEs. However, with the states of UEs varying, time misalignments occur inevitably, which is commonly detected by the synchronization maintenance procedure and eliminated by stepping back to the initialization operation. One of the essential parts of the processes mentioned above is the TO estimation, which is the point we focus on in this paper.
Methods of TO estimation in the OFDM have been widely investigated in \cite{c24,c25,c27,c28,c29,c30,c31,c32,c34}. However, some TO estimators developed for OFDM may not be applicable to UFMC, e.g., the additional sub-band filters in UFMC will destroy the self-correlation and constant modulus properties of Constant Amplitude Zero Auto Correlation sequence, which is commonly used in OFDM TO estimation algorithm. Moreover, due to the lack of CP, methods exploiting the CP structures in OFDM become useless in UFMC.

Existing TO estimation methods for UFMC, e.g. \cite{c12,c13,c14,c15,c41}, are originated from that of OFDM. The essence of these methods is to form a specially designed training sequence, which is then used at receivers to estimate TO through finding the peaks of the correlation function curves. Moreover, compared to TO estimating algorithms in OFDM, these methods are modified only according to the different system structure of UFMC, with out considering the potential possibility of multi-user joint estimation, as well as the underlying multi-user joint sparsity. On the other hand, the precision of these on-grid algorithms is limited in particular, for the nature that TO estimation is essentially an off-grid problem.

Motivated by the limitations of above-mentioned algorithms, in this paper, we devise a joint timing synchronization and channel estimation scheme for UFMC-based systems to improve the TO estimation accuracy and reduce synchronization overhead during the maintenance phase, freeing the system from continuous signalling exchanging and unnecessary reinitialization operation. Here, assisted by the training pilots for channel estimation, the joint estimator of TO and channel is formulated as a convex optimization problem under the atomic norm minimization (ANM) framework that exploits the multi-user-wise joint sparsity in the continuous TO. Simulation results are provided to illustrate the superior performances of the proposed method over existing solutions.

\section{Signal Model}

\begin{figure}[htbp]
		\centering
		\includegraphics[scale=0.40]{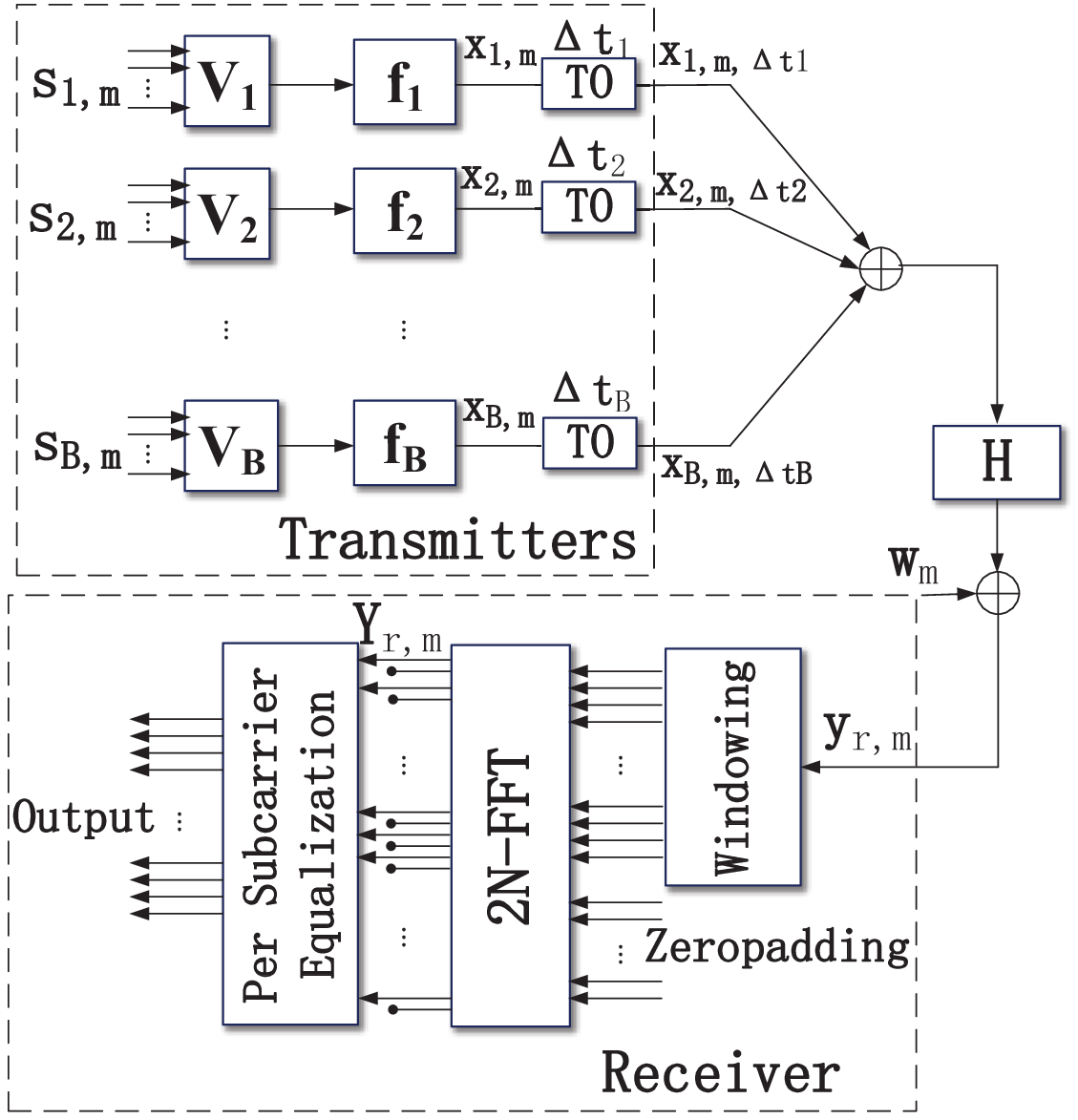}
		\caption{Multi-user uplink UFMC system model.}
		\label{fig:UFMC}
	\end{figure}

We consider a multi-user UFMC uplink system, as shown in Fig. \ref{fig:UFMC}, where the system bandwidth is divided into $B$ sub-bands and each sub-band comprises of $n_s$ consecutive sub-carriers. Without loss of generality, we assume $B$ sub-bands are allocated to the $B$ UEs.

At the transmitter, the $m$th data symbol vector on the $i$th sub-band $\mathbf{s}_{i,m}=[s_{i,m}(0),...,s_{i,m}(n_s-1)] \in \mathcal{M}^{n_s}$ with $\mathcal{M}$ being the symbol constellation set is first processed by a partial $N$-point Inverse Discrete Fourier Transform (IDFT) $\mathbf{V}_i$. We denote the IDFT output of the $m$th symbol of the $i$th UE as $s_{i,m}(t)$, i.e.,
\begin{equation}
s_{i,m}(t)=\sum\limits_{k=(i-1)n_s}^{in_s-1}s_{i,m}(k-(i-1)n_s)e^{j2\pi k\frac{t-t_s}{T'_s}},
\end{equation}
where $i=1,...,B$, $m=1,...,M$, $t\in [t_s,t_s+T'_s]$, $t_s$ is the symbol starting time, $T'_s=NT_s$ denotes the symbol duration, with $T_s$ representing the sampling interval, and $N \geq n_sB$. The signal $s_{i,m}(t)$ is then filtered by an Finite Impulse Response (FIR) filter  $f_i(t)$, $t\in[0,LT_s]$, which is employed in the UFMC to deal with the inter-block interference (IBI). Thus the $m$th transmitted signal of the $i$th UE is given by
\begin{equation}
x_{i,m}(t)=s_{i,m}(t)*f_i(t),
\label{eq:x_i}\end{equation}
where $*$ denotes the convolution operation, and $t\in [t_s,t_s+T'_s+LT_s]$.

The channel is block-fading, i.e., the channel keeps invariant within a frame duration of $M$ symbols. The corresponding channel impulse response from the $i$th UE to the BS is denoted as $\tilde{h}_i(t)$, $t\in[0,L_i]$, $i=1,...,B$. The TO induced by inaccurate detection of symbol starting time between the $i$th UE and the BS is represented as $\Delta t_i = (\bar{t}_i - t_s)/{T_s} \in \mathbb{R}$, where $\bar{t}_i$ is the actual symbol staring time, and $t_s$ denotes the detection starting time. Moreover, the $m$th transmitted symbol with TO of $\Delta t_i$ is denoted as $\tilde{x}_{i,m,\Delta t_i}(t)=x_{i,m}(t-\Delta t_iT_s)$, $t\in [t_s,t_s+T'_s+LT_s]$. Thus the received signal at the BS corresponding to the $m$th symbol from the $B$ UEs is given by

\begin{equation}
\begin{split}
\tilde{y}_{m}(t)&=\sum\limits_{i=1}^{B}[\tilde{h}_i(t)*\tilde{x}_{i,m,\Delta t_i}(t)]+w_m(t),\\
t&\in [t_s,t_s+T'_s+LT_s+L_{max}], m = 0,...,M,
\label{eq:yr1}
\end{split}
\end{equation}
where $L_{max}=\max\limits_{i=1,...,B}\{L_i\}$ and $w_m(t)\sim \mathcal{N}(0,\sigma^2)$ is the Gaussian white noise.\\

At the receiver, the $(T'_s+LT_s+L_{max})$-length signal $\tilde{y}_{m}(t)$ is firstly sampled at the sampling interval of $T_s$ to obtain $(N+L-1)$ samples, i.e., $\mathbf{y}_{m}=[y_{m}(0),...,y_{m}(N+L-2)]^T\in \mathbb{C}^{(N+L-1)\times 1}$, where $y_{m}(n)=y_{m}(t_s+nT_s)$, and then $\mathbf{y}_{m}$ is zero-padded and processed by a $2N$-point FFT to obtain
\begin{equation}
\begin{split}
Y_{m}(k)
&=\sum_{n=0}^{N+L-2}y_{m}(n)e^{-j2\pi kn/2N}\\
&=\sum\limits_{i=1}^{B}H_i(k)X_{i,m,\Delta t_i}(k)+W_m(k), k=0,...,2N-1,
\label{eq:Yr2}
\end{split}
\end{equation}
where
\begin{equation}
\begin{aligned}
H_i(k)&=\sum_{l=0}^{L_i-1}h_i(l)e^{-j2\pi kl/2N}\\
X_{i,m,\Delta t_i}(k)&=\sum_{n=0}^{N+L-2}x_{i,m,\Delta t_i}(n)e^{-j2\pi kn/2N},\label{eq:fft}
\end{aligned}
\end{equation}
$\{h_i(l)\}_{l=0,...,\lfloor\frac{L_i}{T_s}\rfloor-1}$ and $\{x_{i,m,\Delta t_i}(n)\}_{n=0,...,N+L-2}$ are the results of sampling $\tilde{h}_i(t)$ and $\tilde{x}_{i,m,\Delta t_i}(t)$ with interval $T_s$, respectively. $W_m(k)\sim \mathcal{N}(0,\sigma^2)$ is the corresponding Gaussian white noise in frequency domain.\\

For simplicity, the sub-carrier channels that belong to the same sub-band are assumed approximately equal, i.e., $H_i(k)=H_i,\forall k$. 
Denoting $\mathbf{Y}_{m} = [Y_{m}(0),Y_{m}(1)...,Y_{m}(2N-1)]$, $\mathbf{X}_{i,m,\Delta t_i} = [X_{i,m,\Delta t_i}(0),...,X_{i,m,\Delta t_i}(2N-1)]$ and $\mathbf{W}_{m} = [W_{m}(0),...,W_{m}(2N-1)]$, we then have
\begin{equation}
\mathbf{Y}_{m}=\sum\limits_{i=1}^{B}H_i\mathbf{X}_{i,m,\Delta t_i}+\mathbf{W}_m \in \mathbb{C}^{2N\times 1}.
\label{eq:yrt}
\end{equation}

\section{Joint TO and Channel Estimation using ANM}

\subsection{TO Interference Analysis}
Before formulating the joint estimation problem, we first analyze the TO interference. For simplicity, we take repetitive pilot sequence for all sub-bands as an example, i.e. $x_{i,m\pm 1}(t) = x_{i,m}(t), \forall i, m$, for both cases of $\Delta t_i >0$ and $\Delta t_i <0$.

\begin{figure}[htbp]
	
	\centering
	\includegraphics[width=3in]{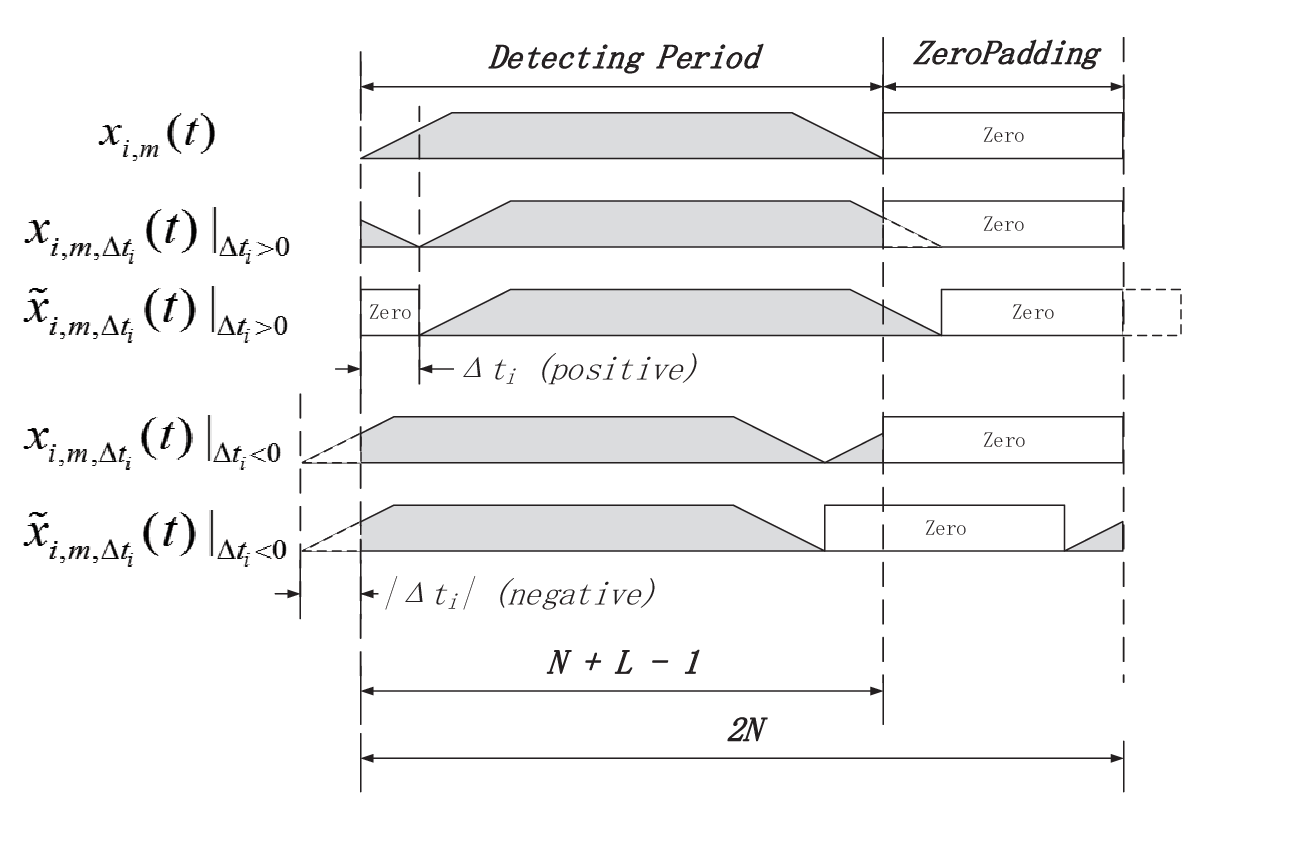}
	
	\centering
	\caption{Timing offset on the $m$th symbol}
	\label{fig:timing offset}
\end{figure}

\textbf{a)} $\Delta t_i >0$: As shown in Fig. \ref{fig:timing offset}, due to the repetitive pilot sequence, the TO leads to the circular shift of $x_{i,m}(t)$ in the detection period. Denoting $\mathrm{CS}[\cdot]_a^U$ as a circular shifting operator, where $U$ is the period, $|a|$ is the shifting length, and the sign of $a$ indicates the shifting direction. Letting $T'_s+LT_s-T_s=T_U$ as the length of a UFMC symbol, thus we have $x_{i,m,\Delta t_i}(t)=\mathrm{CS}[x_{i,m}(t)]_{\Delta t_iT_s}^{T_U}$. Zero padding after sampling at interval $T_s$ is equivalent to sampling the continuous-time signal which is given by (for simplicity we set $t_s=0$)
\begin{equation}
\bar{x}_{i,m,\Delta t_i}(t)=\left\lbrace \begin{split}
&x_{i,m,\Delta t_i}(t),~~~~~~~~0\leq t\leq T_U,\\
&0,~~~~~~~~~~~~~~T_U<t\leq 2NT_s.\end{split}\right.
\end{equation}
Denote $\tilde{x}_{i,m,\Delta t_i}(t)=\mathrm{CS}[x_{i,m}(t)]_{\Delta t_iT_s}^{2NT_s}$ as shown in Fig. \ref{fig:timing offset}. Then we can write $\bar{x}_{i,m,\Delta t_i}(t)$ as 
\begin{equation}
\bar{x}_{i,m,\Delta t_i}(t)=\tilde{x}_{i,m,\Delta t_i}(t)+I_i^+(t),\label{eq:barx}
\end{equation}
where 
\begin{equation}\begin{split}
I_i^+(t)&=\bar{x}_{i,m,\Delta t_i}(t)-\tilde{x}_{i,m,\Delta t_i}(t)\\
&=\left\lbrace \begin{split}
&x_{i,m}(t-\Delta t_iT_s+T_U),~~~~~~~0\leq t < \Delta t_iT_s,\\
&0,~~~~~~~~~~~~~~~~~~~~~~~\Delta t_iT_s+T_U\leq t< T_U,\\
&-x_{i,m}(t-\Delta t_iT_s),~T_U\leq t < \Delta t_iT_s+T_U,\\
&0,~~~~~~~~~~~~~~~~~~~~~~\Delta t_iT_s+T_U\leq t\leq 2T'_s.
\end{split}\right.
\end{split}\end{equation}
Denoting $\mathrm{FFT}[\cdot]$ as the operation of sampling at the interval $T_s$ and followed by 2N-point Fast Fourier Transformation (FFT). On the basis of \eqref{eq:barx}, we have
\begin{equation}
\mathbf{X}_{i,m,\Delta t_i}=\mathrm{FFT}[\bar{x}_{i,m,\Delta t_i}(t)]=\tilde{\mathbf{X}}_{i,m,\Delta t_i}+\mathbf{I}_i^+,
\end{equation}
where $\tilde{\mathbf{X}}_{i,m,\Delta t_i}=\mathrm{FFT}[\tilde{x}_{i,m,\Delta t_i}(t)]$, $\mathbf{I}_i^+=\mathrm{FFT}[I_i^+(t)]$. Then by the property of Fourier transform, we have $\tilde{\mathbf{X}}_{i,m,\Delta t_i}=\mathbf{X}_{i,m}\odot\mathbf{e}_{\Delta t_i}$, with $\mathbf{e}_{\Delta t_i}=[1,e^{-j2\pi \cdot 1 \cdot \Delta t_i/2N},...,e^{-j2\pi \cdot (2N-1) \cdot \Delta t_i/2N}]^T \in\mathbb{C}^{2N\times1}$ and $\mathbf{X}_{i,m}=\mathrm{FFT}[x_{i,m}(t)]$. Therefore, we get
\begin{equation}
\mathbf{X}_{i,m,\Delta t_i}=\mathbf{X}_{i,m}\odot\mathbf{e}_{\Delta t_i}+\mathbf{I}_i^+.\label{eq:y+t0}
\end{equation}


\textbf{b)} $\Delta t_i <0$: Similarly, the derivation of $\mathbf{X}_{i,m,\Delta t_i}$ can be given by
\begin{equation}
\mathbf{X}_{i,m,\Delta t_i}=\mathbf{X}_{i,m}\odot\mathbf{e}_{\Delta t_i}+\mathbf{I}_i^-,\label{eq:y-t0}
\end{equation}
where $\mathbf{I}_i^-=\mathrm{FFT}[I_i^-(t)]$ with
\begin{equation}
I_i^-(t)=\left\lbrace \begin{split}
&0,~~~~~~~~~~~~~~~~~~~~~~~0\leq t< T_U+\Delta t_iT_s,\\
&x_{i,m}(t-\Delta t_iT_s-T_U),\\
&~~~~~~~~~~~~~~~~~~~~~~~~T_U+\Delta t_iT_s\leq t < T_U,\\
&0,~~~~~~~~~~~~~~~~~~~~T_U\leq t< 2T'_s+\Delta t_iT_s,\\
&-x_{i,m}(t-\Delta t_iT_s-2T'_s),\\
&~~~~~~~~~~~~~~~~~~~~~~2T'_s+\Delta t_iT_s\leq t \leq 2T'_s.
\end{split}\right.
\end{equation}

The first terms in \eqref{eq:y+t0} and \eqref{eq:y-t0}, i.e. $\mathbf{X}_{i,m}\odot\mathbf{e}_{\Delta t_i}$, can be used to estimate TO and channels, while the terms $\mathbf{I}_i^+$ and $\mathbf{I}_i^-$ are interference.

Denoting $\eta_i^+=\frac{E\left\lbrace |\mathbf{X}_{i,m}\odot\mathbf{e}_{\Delta t_i}|^2\right\rbrace}{E \left\lbrace |\mathbf{I}_i^+|^2\right.\rbrace}$ and $\eta_i^-=\frac{E\left\lbrace |\mathbf{X}_{i,m}\odot\mathbf{e}_{\Delta t_i}|^2\right\rbrace}{E \left\lbrace |\mathbf{I}_i^-|^2\right.\rbrace}$ as the power ratios of useful signal to TO-interference for the cases of $\Delta t_i >0$ and $\Delta t_i <0$, respectively.

\begin{figure}[htbp]
	
	\centering
	\includegraphics[width=3in]{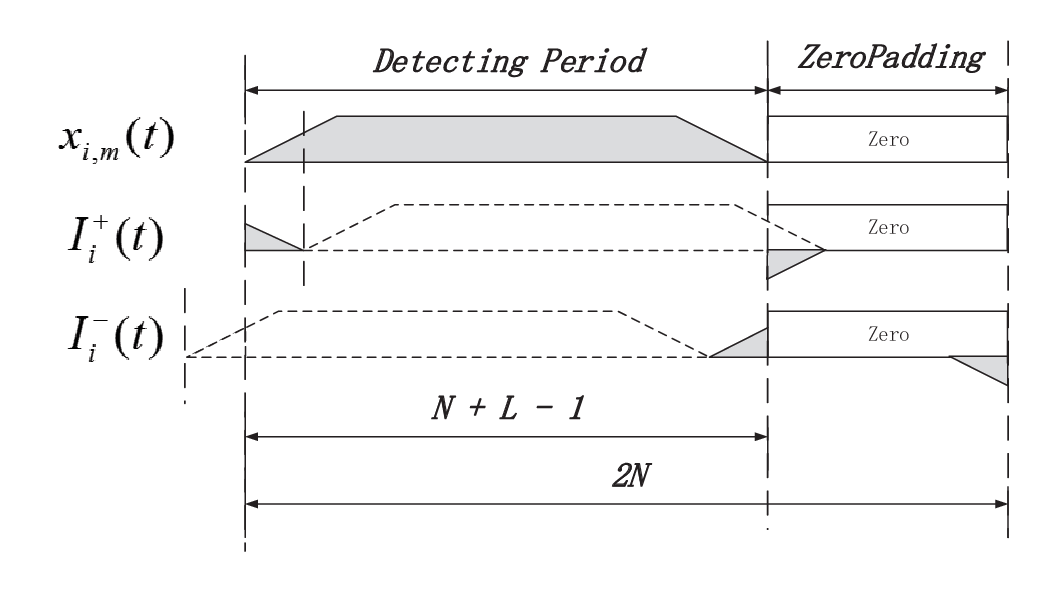}
	\centering
	\caption{Interfering part of the received signal}
	\label{fig:It}
\end{figure}

As shown in Fig. \ref{fig:It}, $I_i^+(t)$ and $I_i^-(t)$ are the combinations of the head and tail parts of the transmitted signal. Benefiting from sub-band filters employed in UFMC, both two ends of the transmitted symbol occupy much less energy than the central part of it, i.e., $E \left\lbrace |\mathbf{I}_i^+|^2\right.\rbrace\ll E\left\lbrace |\mathbf{X}_{i,m}\odot\mathbf{e}_{\Delta t_i}|^2\right\rbrace$, $E \left\lbrace |\mathbf{I}_i^-|^2\right.\rbrace\ll E\left\lbrace |\mathbf{X}_{i,m}\odot\mathbf{e}_{\Delta t_i}|^2\right\rbrace$. Therefore, $\eta_i^+$ and $\eta_i^-$ are usually high enough. In other words, the interference caused by $\mathbf{I}_i^+$ and $\mathbf{I}_i^-$ is even lower than that of AWGN in practical communication systems.

Thus, ignoring the terms $\mathbf{I}_i^+$ and $\mathbf{I}_i^-$, we get the receiver 2N-point FFT output based on \eqref{eq:yrt}, \eqref{eq:y+t0} and \eqref{eq:y-t0} as
\begin{equation}
\mathbf{Y}_{m}=\sum\limits_{i=1}^BH_i\mathbf{X}_{i,m}\odot\mathbf{e}_{\Delta t_i}+\mathbf{W}_m.
\end{equation}

To facilitate our analysis development, we assume the pilot symbols remain the same for different sub-bands, i.e., $\mathbf{X}_{i,m} = \mathbf{X}_{m},\forall i$. Finally we have
\begin{equation}
\mathbf{Y}_{m}=\mathbf{X}_m\odot\sum\limits_{i=1}^BH_i\mathbf{e}_{\Delta t_i}+\mathbf{W}_m.
\end{equation}


\subsection{Joint Estimator}

Our target is to jointly estimate $\{H_i\}_{i=1}^B$ and $\{\mathbf{e}_{\Delta t_i}\}_{i=1}^B$ from $\mathbf{Y}_{m}$, where $\Delta t_i$ is a continuous-time variable. To solve the off-grid problem, we use the atomic norm to enforce the sparsity of $\{\mathbf{e}_{\Delta t_i}\}_{i=1}^B$.\\


Denote $\mathbf{e}(\tau_i)=[1,e^{j2\pi\tau_i},...,e^{j2\pi (2N-1)\tau_i}]^T$, where $\tau_i\triangleq -\Delta t_i/2N$,  $\mathbf{g}=\sum_{i=1}^{B}H_i\mathbf{e}(\tau_i)\in\mathbb{C}^{2N\times 1}$, then we have $\mathbf{Y}_{m}=\mathbf{X}_{m}\odot\mathbf{g}+\mathbf{W}_m$. The atomic norm for $\mathbf{g}\in\mathbb{C}^{2N\times1}$ is
\begin{equation}
\Vert\mathbf{g}\Vert_{\mathcal{A}}=\inf_{\tau_i\in [-\frac{1}{2},\frac{1}{2}]}\left\lbrace \sum_{i=1}^{B}|H_i|:\mathbf{g}=\sum_{i=1}^{B}H_i\mathbf{e}(\tau_i)\right\rbrace.\label{eq:atom}
\end{equation}


Then we formulate the joint estimator based on the following optimization
\begin{equation}
\hat{\mathbf{g}}=\mathop{\arg\min}_{\mathbf{g}\in\mathbb{C}^{2N\times1}, \tau_i\in [-\frac{1}{2},\frac{1}{2}]} \|\mathbf{g}\|_{\mathcal{A}}+\lambda\|\mathbf{Y}_{m}-\mathbf{X}_{m}\odot\mathbf{g}\|_2^2,
\end{equation}
where $\lambda >0$ is the weight factor. In practice, we set $\lambda\simeq\sigma\sqrt{2N\log(2N)}$.\\
Based on the convex equivalent formulation of atomic norm, we can get the following semidefinite program (SDP):
\begin{equation}
\left\{ \begin{aligned}
\mathop{\min}_{\begin{subarray}{c}\mathbf{T}\in\mathbb{C}^{2N\times 2N},\\ \mathbf{g}\in\mathbb{C}^{2N\times1},t\in\mathbb{R}^+\end{subarray}} &\frac{1}{4N}\mathrm{Tr}(\mathbf{T})+\frac{t}{2}+\lambda\|\mathbf{Y}_{m}-\mathbf{X}_{m}\odot\mathbf{g}\|_2^2
\\s.t.~~~~~~&\left[ \begin{matrix}\mathbf{T} & \mathbf{g}\\
\mathbf{g}^H & t\end{matrix}\right] \succeq 0,
\end{aligned}\right
.\label{eq:sdp}\end{equation}
where $\mathrm{Tr}(\cdot)$ denotes the trace, $t=\sum_{i=1}^{B}|H_i|$, $\mathbf{T}=\sum_{i=1}^{B} |H_{i}|^2\mathbf{e}_{\tau_i}\mathbf{e}_{\tau_i}^{H}\in\mathbb{C}^{2N\times2N}$is a Toeplitz matrix.

The above problem is convex, so it can be solved efficiently using a convex solver.

We denote the solutions to \eqref{eq:sdp} as $\hat{\mathbf{g}}$ and $\mathbf{T}$. Notice that the resulted $\mathbf{T}$ meets the form of Vandemonde Decomposition. 
Thus we choose the matrix pencil method to extract $\{\hat{\tau}_i\}_{i=1}^B$ from the matrix $\mathbf{T}$ as follow\cite{c35}.

Since $\mathbf{T}$ is positive semi-definite with $\mathrm{rank}(\mathbf{T})=B$, there exists $\mathbf{D}\in\mathbb{C}^{2N\times B}$ such that $\mathbf{T}=\mathbf{DD}^H$. Write $\mathbf{D}$ as $\mathbf{D}=[\mathbf{d}_0^H,...,\mathbf{d}_{2N-1}^H]^H$ with $\mathbf{d}_n\in\mathbb{C}^{1\times B},n=0,...,2N-1$. Define the upper submatrix $\mathbf{D}_U=[\mathbf{d}_0^H,...,\mathbf{d}_{2N-2}^H]^H$ and the lower submatrix $\mathbf{D}_L=[\mathbf{d}_1^H,...,\mathbf{d}_{2N-1}^H]^H$. 
We consider the matrix pencil $(\mathbf{D}_U^H\mathbf{D}_L,\mathbf{D}_U^H\mathbf{D}_U)$. By solving a generalized eigenproblem, we get the eigenvalues of $(\mathbf{D}_U^H\mathbf{D}_L,\mathbf{D}_U^H\mathbf{D}_U)$ as $\{d_i\}_{i=1}^B$, which holds $d_i=e^{j2\pi\tau_i},i=1,...,B$. Thus $\{\hat{\tau}_i\}_{i=1}^B$ is obtained, and $\{\Delta\hat{t}_i\}_{i=1}^B$ and $\{\hat{\mathbf{e}}(\tau_i)\}_{i=1}^B$ can be further calculated accordingly. 
We denote $\hat{\mathbf{E}}=[\hat{\mathbf{e}}(\tau_1),...,\hat{\mathbf{e}}(\tau_B)]$ and $\mathbf{h}=[H_1,...,H_B]^H$. Obviously it holds that $\hat{\mathbf{E}}\mathbf{h}=\hat{\mathbf{g}}$.
Using the Least Square method, $\hat{\mathbf{h}}$ can be calculated by
\begin{equation}
\hat{\mathbf{h}}=(\hat{\mathbf{E}}^H\hat{\mathbf{E}})^{-1}\hat{\mathbf{E}}^H\hat{\mathbf{g}},
\end{equation}
where $\hat{\mathbf{h}}\triangleq [\hat{H}_1,...,\hat{H}_B]^H$ indicates the estimates of channel coefficients.

For the proposed ANM-based estimator, $\left[ \begin{matrix}\mathbf{T} & \mathbf{g}\\
\mathbf{g}^H & t\end{matrix}\right]$ in \eqref{eq:sdp} is of size $(2N+1)^2$, thus its complexity is $\mathcal{O}(N^3)$.


After obtaining the estimated $\{\Delta\hat{t}_i\}_{i=1}^B$, the BS decides whether to send a new Timing Advance message to a certain UE by comparing $\Delta\hat{t}_i$ with the given threshold. Thus, the BS is able to adjust the timing of a UE when needed, instead of keeping strict synchronization by continuous signaling exchange in each period. 

\section{SIMULATION RESULT}
\begin{figure}[htbp]
	\centering
	\subfigure[NMSE of timing offsets]{
		\begin{minipage}[t]{0.5\linewidth}
			\centering
			\includegraphics[width=1.6in,height=1.2in]{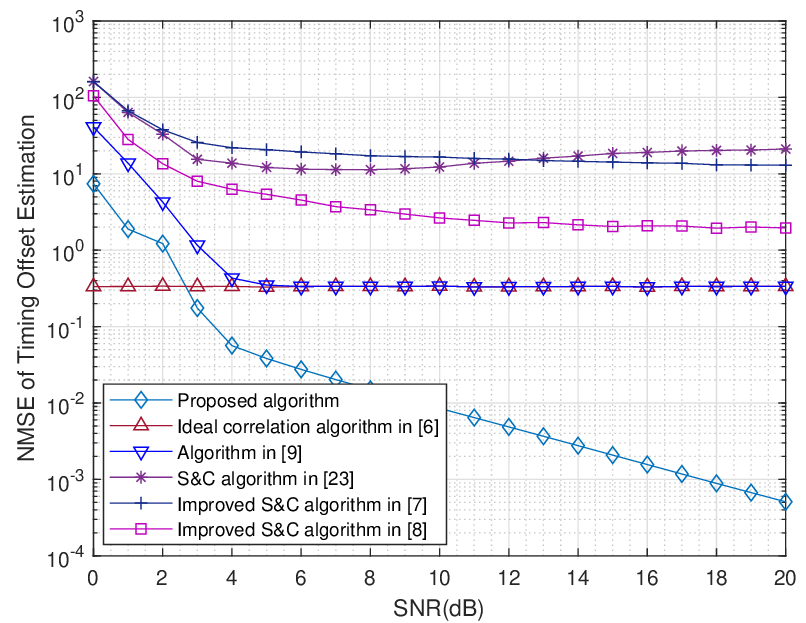}
			\label{fig:nmse_t}
		\end{minipage}%
	}%
	\subfigure[Bit error rate]{
		\begin{minipage}[t]{0.5\linewidth}
			\centering
			\includegraphics[width=1.6in,height=1.2in]{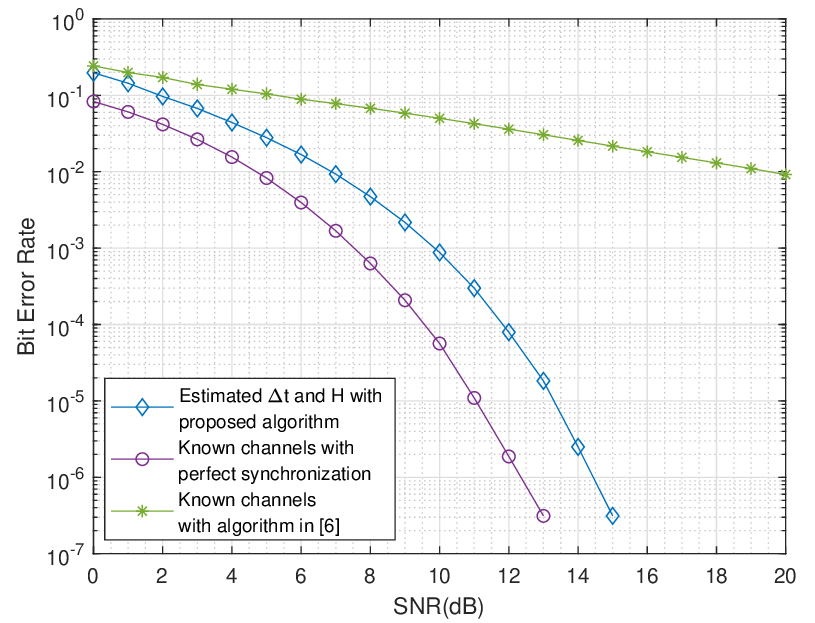}
			\label{fig:ber}
		\end{minipage}%
	}%
	\centering
	\caption{System performance using the proposed algorithm: (a) NMSE performance comparison of TO estimation between proposed algorithm and existing methods; (b) BER performance comparison between proposed algorithm and system with perfect synchronization}
\end{figure}
In this section, we use simulations to illustrate the performance of the proposed algorithm for both the $\Delta t_i > 0$ and $\Delta t_i < 0$ cases. The number of the IDFT points is set as $N=64$; the number of sub-bands is $B = 2$, which are allocated to two different UEs. Following the setting of \cite{c38}, each sub-band occupies $n_s = 16$ subcarriers. Transmitters using the length $L$ Dolph-Chebyshev filter whose Fourier transform side-lobe magnitude is $\alpha$ dB below the main-lobe magnitude, where $L=6,\alpha=120$.
The number of symbols is $M = 10^5$ and QPSK modulation is employed. The channel $\{H_i\}_{i=1}^B$ are randomly generated according to distribution $\mathcal{CN}(0,1)$. The TO $\{\Delta t_i\}_{i=1}^B$ are randomly generated according to distribution $\mathcal{U}(-L,L)$ where $L$ is the length of transmitting filter. The signal-to-noise ratio (SNR) is defined as $\frac{P}{2N\sigma^2}$ with $P$ denoting the average transmission power. We use the NMSE defined as $10 \log_{10}(\mathbb{E}[\frac{1}{B}\sum_{i=1}^B(\Delta\hat{t}_i-\Delta t_i)^2])$ to evaluate the performance of TO estimation, and the BER defined as $\mathbb{E}[\sum_{m=1}^M\|\mathbf{s}_m-\hat{\mathbf{s}}_m\|_0/(M\times N_s)]$ with $N_s$ denoting the number of bits in $\mathbf{s}_m$, to evaluate the performance of symbol demodulation.

For the first experiment, as shown in Fig. \ref{fig:nmse_t}, to compare with the proposed estimator, we consider five methods, i.e., the ideal correlation algorithm\cite{c12}, algorithm in \cite{c15}, and the S\&C algorithm\cite{c41} with its two improved methods\cite{c13,c14}. All these algorithms are based on training sequences with specially designed structures. The algorithm in \cite{c12} transmits several consecutive Zadoff-Chu sequences, and at the receiver, correlates the received signal with the Zadoff-Chu sequence itself, followed by the shifting and superposition, to get the peak index as the TO estimation. The simulation result shows its higher precision than other existing methods, and the accuracy keeps almost unchanged under various SNR. It is regarded as the ideal case of correlation algorithms due to ignoring the overhead and sub-band filters. Similarly, algorithm in \cite{c15} also designs the output signal of sub-band filters directly, to structure a desired training sequence. Because of the reduced number of using symbols, the accuracy of algorithm in \cite{c15} declines in case of low SNR, e.g., $SNR<4dB$. The above mentioned two methods are theoretically able to use in any systems including UFMC, however, there remain problems in physical implementation. The S\&C algorithm\cite{c41} and its modifications\cite{c13,c14} are considered more practical, but suffer low precision due to the plateau formed at the peak of correlation function curves. The plateau makes the desired peak index unable to be locked on accurately. Although the improved algorithms\cite{c13,c14} reduce the plateau-like effect by various means, the estimation precision is still unsatisfactory. On the contrast, the proposed algorithm performs a much higher precision than almost all algorithms simulated above, due to utilizing the underlying multi-user-wise joint sparsity and the off-grid nature of the algorithm. Only the ideal correlation algorithm\cite{c12} keeps advantages in case of low SNR, e.g., $SNR<3dB$. 

For the second experiment, as shown in Fig. \ref{fig:ber}, to verify the statements of the proposed algorithm working in UFMC system, we perform comparison of the BER curves between algorithm in \cite{c15}, the proposed algorithm and the ideal condition which is set as known channel with perfect synchronization. It is obvious that the proposed algorithm performs much lower BER than the algorithm in \cite{c15} (with known channels). Compared to the ideal condition curve, BER of the proposed algorithm declines with SNR increasing, and tends to zero at $SNR>15dB$, where the BER of ideal condition of UFMC reaches the same point at $SNR=13dB$.

\section{CONCLUSIONS}
In this paper, we have proposed super-resolution joint To and channel estimators for multi-user uplink UFMC systems. The proposed estimator is based on the ANM that exploits the sparsity in the continuous TO. Simulation results indicate that the proposed algorithm outperform the counterpart TO estimator. Moreover, the proposed joint estimator can also effectively estimate the channel.


\begin{thebibliography}{00}
\bibitem{c39} S. Duan, K. Chen, X. Yu, M. Qian, ``Automatic Multicarrier Waveform Classification via PCA and Convolutional Neural Networks", \textit{IEEE Access}, vol. 6, pp. 51365-51373, Sept. 2018.
\bibitem{c40} J. Wen, J. Hua, W. Lu, Y. Zhang, D. Wang, ``Design of Waveform Shaping Filter in the UFMC System", \textit{IEEE Access}, vol. 6, pp. 32300-32309, May 2018.
\bibitem{c10} S. Han, Y. Sung, Y. H. Lee, ``Filter design for generalized frequency-division multiplexing", \textit{IEEE Trans. Signal Process.}, vol. 65, no. 7, pp. 1644-1659, Apr. 2017.
\bibitem{c11} M. G. Kibria, G. P. Villardi, K. Ishizu, F. Kojima, ``Throughput enhancement of multicarrier cognitive M2M networks: Universal-filtered OFDM systems", \textit{IEEE Internet Things J.}, vol. 3, no. 5, pp. 830-838, Oct. 2016.
\bibitem{c12} X. Wang, F. Schaich, S. ten Brink, ``Channel Estimation and Equalization for 5G Wireless Communication Systems", 2014.
\bibitem{c13} H. Cho, Y. Yan, G. Chang, X. Ma, ``Asynchronous Multi-User Uplink Transmissions for 5G with UFMC Waveform", \textit{Proc. Wireless Commun. Netw. Conf. (WCNC)}, pp. 1-5, mar. 2017.
\bibitem{c14} X. Yu, Z. Zhou, Y. Gao, ``Improved Symbol Timing Synchronization Algorithm for UFMC System", \textit{Computer Engineering}, vol. 44, no. 11, pp. 105-108,114, 2018.
\bibitem{c15} Y. Li, B. Tian, K. Yi, ``An efficient and hybrid timing offset estimation approach for universal-filtered multi-carrier based systems over multipath Rayleigh fading channel", \textit{Digit. Signal Prog.}, vol. 73, pp. 128-134, Feb. 2018.
\bibitem{c24} C. L. Wang, H. C. Wang, ``On joint fine time adjustment and channel estimation for OFDM systems", \textit{IEEE Trans. Wireless Commun.}, vol. 8, no. 10, pp. 4940-4944, Oct. 2009.
\bibitem{c25} M. Tanda, ``Blind symbol-timing and frequency-offset estimation in OFDM systems with real data symbols", \textit{IEEE Trans. Commun.}, vol. 52, no. 10, pp. 1609-1612, Oct. 2004.
\bibitem{c27} Y. Y. Wang, ``Estimation of CFO and STO for an OFDM using general ICI self-cancellation precoding", \textit{Digit. Signal Prog.}, vol. 31, pp. 35-44, Aug. 2014.
\bibitem{c28} C. L. Wang, H. C. Wang, ``A low-complexity joint time synchronization and channel estimation scheme for orthogonal frequency division multiplexing systems", \textit{Proc. IEEE Int. Conf. Commun. (ICC)}, pp. 5670-5675, Jun. 2006.
\bibitem{c29} B. Park, H. Cheon, C. Kang, D. Hong, ``A novel timing estimation method for OFDM systems", \textit{IEEE Commun. Lett.}, vol. 7, no. 5, pp. 239-241, May 2003.
\bibitem{c30} T. M. Schmidl, D. C. Cox, ``Robust frequency and timing synchronization for OFDM", \textit{IEEE Trans. Commun.}, vol. 45, no. 12, pp. 1613-1621, Dec. 1997.
\bibitem{c31} H. Minn, M. Zeng, V. K. Bhargava, ``On timing offset estimation for OFDM systems", \textit{IEEE Commun. Lett.}, vol. 4, no. 7, pp. 242-244, Jul. 2000.
\bibitem{c32} A. B. Awoseyila, C. Kasparis, B. G. Evans, ``Improved preamble-aided timing estimation for OFDM systems", \textit{IEEE Commun. Lett.}, vol. 12, no. 11, pp. 825-827, Nov. 2008.
\bibitem{c34} H. Abdzadeh-Ziabari, M. G. Shayesteh, ``A novel preamble-based frame timing estimator for OFDM systems", \textit{IEEE Commun. Lett.}, vol. 16, no. 7, pp. 1121-1124, Jul. 2012.
\bibitem{c35} Z. Yang, L. Xie, P. Stoica, ``Vandermonde decomposition of multilevel Toeplitz matrices with application to multidimensional super-resolution", \textit{IEEE Trans. Inf. Theory}, vol. 62, no. 6, pp. 3685-3701, Jun. 2016.
\bibitem{c38} X. Chen, L. Wu, Z. Zhang, J. Dang, J. Wang, ``Adaptive Modulation and Filter Configuration in Universal Filtered Multi-Carrier Systems", \textit{IEEE Trans. Wireless Commun.}, vol. 17, no. 3, pp. 1869-1881, Mar. 2018.
\bibitem{c41} T. Schmidl, D. Cox, ``Robust frequency and timing synchronization for OFDM", \textit{IEEE Trans. Commun.}, vol. 45, no. 12, pp. 1613-1621, Dec. 1997.
\end{thebibliography}
\end{document}